\documentclass[twocolumn]{aastex63}
\usepackage{amsmath}

\shorttitle{Kepler 90g TTV Analysis}
\shortauthors{Liang et al.}

\graphicspath{{./}{figures/}}
\begin{document}

\title{Kepler-90: Giant transit-timing variations reveal a super-puff}

\correspondingauthor{Uro\v s Seljak}
\email{useljak@berkeley.edu }

\author{Yan Liang}
\affiliation{Department of Physics, University of California, Berkeley, Berkeley, CA 94720,USA}
%\email{yanliang@berkeley.edu}

\author{Jakob Robnik}
\affiliation{Department of Physics, ETH-H\"onggerberg, 8093 Z\"urich, Switzerland}
%\email{jakob.robnik@gmail.com}

\author{Uro\v s Seljak}
\affiliation{Department of Physics, University of California, Berkeley, Berkeley, CA 94720,USA}
\affiliation{Lawrence Berkeley National Laboratory, 1 Cyclotron Road, Berkeley, CA 93720, USA}



\begin{abstract}

Exoplanet Transit Timing Variations (TTVs) caused by gravitational forces between planets can be used to determine planetary masses and orbital parameters.
Most of the observed TTVs are small and sinusoidal in time, leading to degeneracies between the masses and orbital parameters.
Here we report a TTV analysis of Kepler-90g and Kepler-90h, which exhibit large TTVs up to 25 hours. 
With optimization, we find a unique solution which allows us to constrain all of the orbital parameters.
The best fit masses for Kepler-90g and 90h are $15.0^{+0.9}_{-0.8}$ $M_{\bigoplus}$ (Earth mass) and $203^{+5}_{-5}M_{\bigoplus}$, respectively, with Kepler-90g having an unusually low apparent density of $0.15\pm 0.05\, {\rm g\,cm^{-3}}$. The uniqueness of orbital parameter solution enables a long-term dynamical integration, which reveals that although their periods are close to 2:3 orbital resonance, they are not locked in resonance, and the configuration is stable over billions of years. The dynamical history of the system suggests that planet interactions are able to raise the eccentricities and break the resonant lock after the initial formation.

\end{abstract}



\section{Introduction} \label{sec:intro}

The Kepler mission has detected thousands of exoplanets via the transit method, in which the brightness of stars is measured precisely and continuously, seeking evidence for periodic dimming events due to eclipsing planets \citep{Borucki2010}. A small percentage of the discovered planets exhibit transit timing variations (TTVs), which can be an indicator of gravitational interaction with other (possibly non-transiting) objects, and have been proposed as a powerful probe of planet masses and eccentricities \citep{Agol05,Holman05}.
TTVs have already been used to detect non-transiting planets \citep{k159,nesvorny2012detection} and to determine planet masses and eccentricities \citep{hadden2017kepler,lithwick2012extracting}.
Most of the existing analyses have been on shorter period planets (under 100 days) with small amplitude TTVs, exhibiting a characteristic sinusoidal structure: these are easy to identify in the data but require many transits. A sinusoidal signal can determine only a limited number of parameters and cannot break the degeneracy between the mass and eccentricity \citep{lithwick2012extracting}. In the purely sinusoidal case, only three independent measurements can be extracted from the observed TTVs, while there are seven parameters to be determined for each planet.

While exoplanet mass and eccentricity statistics are relatively abundant, characterization of the full orbital parameters is rare, and is even rarer for the long-period planets. The Radial Velocity (RV) is insensitive to the mutual inclinations, and direct imaging can only place a loose bound on the dynamical parameters\citep{wang2018dynamical}. 
Combining RV with TTVs can break some, but not all, of the degeneracies inherent to RV or small TTV alone \citep{Petigura18}. However, orbital parameters can be extremely valuable as they provide important information about the exoplanet formation, migration, and evolution that is not available otherwise \citep{Petigura18,Winn2015}. 

Super-puffs are planets with small masses and large radii, corresponding to a very low density. Most super-puffs are found close to their host star (0.1$\sim$0.3AU) and their masses are typically measured from TTVs \citep{masuda2014very,jontof2014kepler,ofir2014independent}. Two possible explanations of the low inferred densities are proposed: the absorption from a dusty atmosphere (e.g. \cite{wang2019dusty}) and the effect of exo-rings (see e.g. \citep{piro2020exploring} for a discussion).
The atmosphere accretion and migration mechanisms are discussed by \citep{ChoksiChiang2020}. In this paper we 
will define a super-puff as a planet with an apparent
density less than $0.3{\rm g cm^{-3}}$, regardless of 
the origin. 

The Kepler-90 system has more known transiting planets than almost any other system. Among its seven confirmed planets, the inner five have orbital periods ranging from 7 to 125 days, while 
the outer two are 90g and 90h with orbital periods of 210.5 days and 331.6 days, respectively \citep{kepler90}. 
Based on the 6 recorded transits of 90g and 3 transits of 90h during the four years of Kepler observations, both planets show significant TTVs. Most notably, the sixth observed transit of 90g exhibits a 25-hour anomaly, the largest TTV found to date.

We analyzed the TTVs using a likelihood analysis pipeline 
developed by \cite{FourierGP}: this analysis 
models the non-Gaussian noise of Kepler 
data, and performs a Fourier based Gaussian process analysis, fitting star variability parameters 
with optimal prior power spectrum jointly with the planet parameters. 
The resulting outputs are TTVs and their associated uncertainties. We use the long cadence data (30 minute sampling), supplementing with the short cadence data when available (transits 4-6 of Kepler-90g and transit 3 of Kepler-90h). 
We use the same analysis to output 
the timing duration variability (TDV) of each event and the associated error, for a total of 18 data points, 9 TTVs and 9 TDVs.

\begin{figure*}[htb!]
\centering
\includegraphics[width=0.8\linewidth]{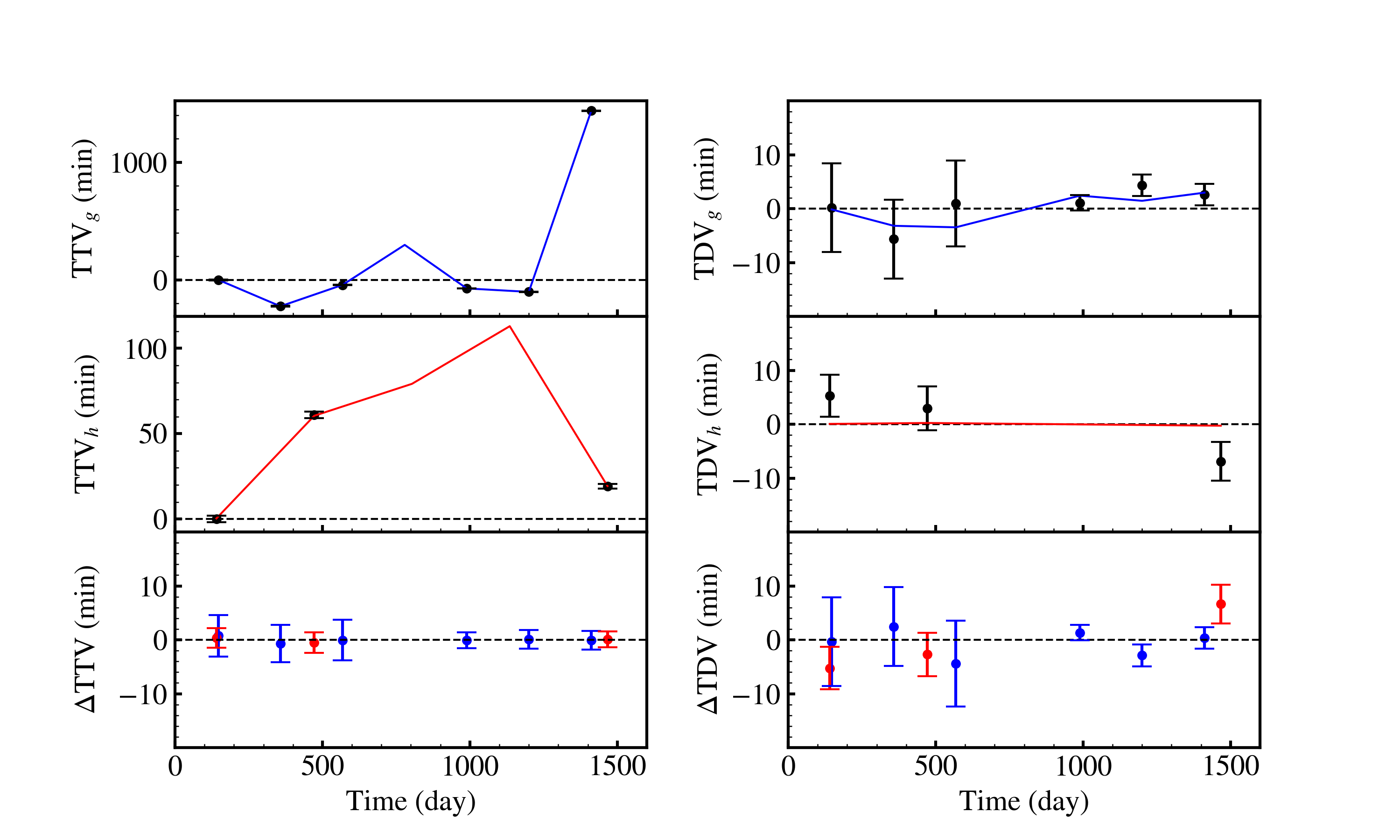} 
\label{fig:TTV}
\caption{Left column: The top, middle and bottom panels show Kepler 90g TTV, Kepler 90h TTV, and TTV residuals of the best two-planet model, respectively. The blue solid line in the left panel is the predicted Kepler-90g TTV, and the red solid line is 90h TTV, including the unobserved transits. The black points show the observed TTV. Right column: the same, but for TDV. The total $\chi^2$ of 9.19 is composed of $\chi^2_{\text{TTV}}=0.19$ and $\chi^2_{\text{TDV}}=9.00$, indicating a good fit to TTV and TDV.}
\end{figure*}

\section{Numerical Analysis}

The TTV and TDV observations, the errors, together with the model prediction as a function of the orbital parameters, allow us to form a data likelihood. 
Given the data likelihood of observations, as well as the priors on the orbital parameters, the goal is to determine their posterior distribution. 

Solving this inverse problem is challenging because there are 13 parameters, a relatively small data points (9 TTVs and 9 TDVs), and the expensive-to-evaluate and highly nonlinear dynamical model. TTVs can yield useful constraints even with such few transits because TTVs can vary by orders of magnitude even with slight changes in orbital parameters\citep{Rodriguez18}. In contrast, TDVs are mainly sensitive to secular effects and provide more limited information about orbital parameters.

The main challenge is finding the solution: most of the prior space is very far from the solution and the current full dynamical N-body integration models are expensive and difficult to use in inverse problem optimization. Analytic gradients were not available for TTVs until very recently \citep{agol2020refining} and due to the high nonlinearity of the model, the local gradient direction may not be very useful even if it were available. 
A dynamical fit to the sinusoidal small TTV signals usually requires many transits to determine a unique solution. As a consequence, a full dynamical analysis has predominantly been performed on short period inner planets ($P<100$ days) with many transits \citep{hadden2017kepler}. 

Since orbit fitting using the full N-body dynamics is computationally expensive, many works turn to analytic TTV predictions \citep{lithwick2012extracting}, which often suffice for the small TTVs. 
Analytic methods are however not very useful for our TTV analysis: these methods are based on first-order perturbative expansion in terms of eccentricity, characterizing TTVs as sinusoidal signals, while the observed TTV signals are non-sinusoidal (Figure 1). We will argue that this is a result of the strong planetary interactions between Kepler 90g and 90h: 90g has a 25-hour TTV anomaly, 0.5\% of the orbital period, while typical observed planetary TTVs are seconds to minutes. The small errors of individual TTVs, typical of long period transits, can make it easier to observe the departures from the sinusoidal model despite the fewer transits. As a result, the analytic TTV calculations are not accurate enough for the orbit fitting.
However, while the full dynamical analysis is 
difficult, when applied to large TTVs it also provides a unique opportunity to go beyond the sinusoidal analysis and measure the angular parameters that cannot be otherwise constrained. 

While any existing body in the system may perturb the 90g and induce TTVs, a 25-hour excursion is too large to be caused solely by a small object like an asteroid or a moon, or distant objects such as the inner planets \citep{kepler90}. 
The most likely source of this anomaly is the neighboring gas giant, Kepler 90h. Thus, the simplest hypothesis is to model the TTVs with interactions between 90g and 90h, and we first model the TTVs with three bodies: the host star, 90g, and 90h. We verified through numerical experiments that Kepler 90g and 90h does not induce observable TTVs on the interior planets, consistent with the theoretical prediction \citep{Agol05}. Similarly, the inner planets do not produce significant TTVs on 90g and 90h (See Appendix A). It is also recognized in \cite{kepler90} that Kepler 90e/f and 90g/h are dynamically decoupled, so it is safe to ignore the inner planets when studying the 90g/h pair. 

%\section{Fitting Algorithm}
To obtain the model prediction, we use a non-relativistic N-body integration code TTVFast \citep{ttvfast}, as relativistic effects are negligible for such long periods. Given one set of orbital parameters, this program integrates over 1500 days in 0.01-day step-size and detects all the transiting events. Then, we compare the modeled TTVs with the observed signal to evaluate the likelihood of one particular model, and we run an optimization routine to find the set of planetary parameters that maximize the posterior. We describe the details of the optimization procedure in Appendix A.

We obtain the best-fit solution which greatly reduces the TTV residuals from up to 25 hours to less than 1 minute (Figure 1). We searched extensively for additional solutions, by scanning over the parameter space on fine grids (see Appendix A), and by running the \texttt{dynesty} sampler on an uninformative prior (same as the one reported in Table \ref{tab:messier})\citep{speagle2018dynesty}. Both methods converged to the same best solution, and the second best one has an unacceptable $\chi^2 > 100$, so to the best of our knowledge the solution is unique. 

From the MCMC posterior analysis we find 
the best fit masses are $15.0^{+0.9}_{-0.8}$ $M_{\bigoplus}$  and  $203^{+5}_{-5}M_{\bigoplus}$ for 90g and 90h, respectively, consistent with the TTVs in 90g being significantly larger than in 90h. 

The residual $\chi^2$ is 9.19 for 18 data points and is a good fit to the data.
We perform a posterior analysis of the two-planet model, which is numerically challenging due to the highly correlated orbital parameters. We reparameterized the orbital parameters to break the degeneracy and speed up the convergence. Details are described in the Appendix A.
The posterior means and variances were obtained using the MCMC \texttt{emcee} \citep{emcee} package and are shown in Table 1. The posterior plots reveal some remaining strong correlations between the parameters. However, 
because of the extreme sensitivity of TTVs and TDVs to these parameters, the marginalized posteriors are narrow relative to 
the prior for all of them.

\begin{deluxetable*}{lccrr}[htb!]
\tablenum{1}

\tablecaption{Summary of parameter priors and posteriors\label{tab:messier}}
\tablewidth{0pt}
\tablehead{
\colhead{Parameter} & \colhead{90g prior}  & \colhead{90h prior} &
\colhead{90g mean$\, \pm 1 \sigma$} & \colhead{ 90h mean$\,\pm 1 \sigma$} 
}

\startdata
$m (M_{\oplus})$ & $(5,50)$ & $(5,500)$ & $15.0^{+0.9}_{-0.8}$ & $203^{+5}_{-5}$ \\ 
$P$(day) & $(209,212)$& $(330,333)$ & $210.48^{+0.05}_{-0.05}$ & $331.657^{+0.009}_{-0.008}$ \\ 
$e$ & $(0,0.15)$ &$(-0.075,0.075)^1$ & $0.049^{+0.011}_{-0.007}$ & $-0.011^{+0.002}_{-0.003}$ \\ 
$i(^{\circ}) $& $(89.8,90.0)$ &  $(89.8,90.0)$ & $89.92^{+0.03}_{-0.01}$ & $89.927^{+0.011}_{-0.007}$ \\ 
$\Omega(^{\circ}) $  & $0^2$ & $(-180,180)^1$ & $0^2$ & $-2^{+6}_{-6}$ \\ 
$\omega(^{\circ}) $ & $(0,180)$&$(-180,180)^1$ & $90^{+20}_{-20}$ & $-3^{+3}_{-2}$ \\ 
$\omega+M(^{\circ}) $ &$(0,360)$&$(0,360)$  & $198^{+2}_{-2}$ & $297^{+1}_{-1}$ \\ 
\enddata

\tablecomments{The orbital parameters used in the 
 analysis: $m$ is the planetary mass in earth mass unit; $P$ is the orbital period in days; $e$ is eccentricity; $i$  is orbital inclination; $\Omega$ is the mutual inclination; $\omega$ is the argument of the periapsis; $M$ is mean anomaly. All the angles are measured in degrees. Details about the choice of priors can be found in Appendix A. The middle columns show the prior range of Kepler 90g and 90h planetary parameters, respectively.  The right columns contain the mean and $1 \sigma$ errors from the posterior distribution obtained through MCMC sampling. 
 $^1$: the marked 90h parameters are measured from the respective parameter of 90g to improve MCMC sampling if strongly correlated. 
 $^2$: $\Omega$ for 90g is set to 0, defining the $x$ axis.}
\end{deluxetable*}

The near-perfect fit we obtained suggests that there is no need to seek further solutions with a third body. Nevertheless, we describe such a search in the Appendix, confirming that no solutions exist that do not require mutual perturbations of the two planets as the main origin of the TTVs. 

\section{Long-term Stability}

The long-term stability of Kepler 90g and 90h system with a large TTV can be qualitatively understood in terms of the so-called Hill stability, which quantifies how close the two planets can be while still orbiting a star. A sufficient condition for a co-planar two-planet system with initially circular orbits to be stable is $\Delta > 2.4(\mu_1+\mu_2)^{1/3}$,  where $\mu_1,\mu_2$ are the planet to star mass ratios of the inner and outer planet, respectively, and $\Delta = a_2/a_1 -1$, where $a_1, a_2$ refers to the semi-major axis \citep{gladman1993dynamics}. In Kepler 90g and 90h system, $\Delta = 0.35$, and the right-hand side gives 0.19, so the system is Hill stable.

A unique and accurate solution to all planetary parameters enables a quantitative long-term analysis of the orbits.
We ran an N-body integration from today's epoch using the WHFast integrator in\texttt{rebound} package \citep{rebound}. The time span is chosen to be $2$ billion years, the same as the age of the star. We started from the best-fit solution and integrated using a 2-day step size, 1\% of the shortest period in the system. As demonstrated in figure 2, the two-planet orbit remains stable during the course of integration. Due to the strong planetary interaction, some of the orbital parameters are not constant. We explored variations of the orbital parameters by sampling from their posterior and found that they do not significantly affect the long term orbit integration solution. 

We monitor the evolution of the semi-major axis, eccentricity, apsidal angle ($\Delta \omega=\omega_h - \omega_g$), and resonant angle of Kepler 90g and 90h during the 2 billion years integration (Figure 2). The resonant angle is defined as $\phi_{g,h}=3\lambda_h - 2\lambda_g-\varpi_{g,h}$, where $\lambda$ stands for the mean longitude, $\varpi$ is the longitude of the periapsis, and the subscripts denote planet 90g or 90h. If $\phi$ oscillates around zero, it is librating and the system experiences orbital lock. In our case $\phi$ circulates, suggesting that Kepler 90g and 90h are not locked in resonance. \footnote{ It seems that the system temporarily migrates into a 2:3 orbital resonance ($\phi_h$ librates) at $\sim$600 Myr. This is an artifact of the output cadence of data and will be gone if we pick a different cadence.} We also performed a relatively short integration of 1 million years on 100 nearby solutions drawn from the MCMC posterior. All the solutions show circulating $\phi_g, \phi_h$ throughout the integration.

Kepler observations show there is an excess of planet pairs observed just wide of resonances \citep{fabrycky2014architecture,ChoksiChiang2020,lithwick2012resonant}. Kepler 90g and 90h also have a period ratio slightly wide of 2:3. 
This may be explained by the resonant repulsion theory: when the planets are formed in disk,  eccentricity damping drives planet pairs wide of resonance, while they remain in orbital lock \citep{ChoksiChiang2020}.  
This process concludes after the first 10 million years and what happens afterward is unknown. In the Kepler 90g and 90h system, we observe a larger eccentricity than the 1\% upper limit predicted by this theory, and the two planets are not in an orbital lock (Figure 2). The spontaneous migration in and out of the orbital lock we observe in the long 
time integration suggests that planetary interactions may be able to raise the eccentricities and break the resonant lock after the disk dissipates.

The orbital parameter solution also reveals several features that may help preserve the long-term stability, withstanding strong perturbations. The eccentricities are small, 0.045 and 0.035, which means that the orbits are near-circular. However, 0.045 is significantly larger than 0.03 which was assumed to be an upper limit for the long-term stability \citep{kepler90}.

\begin{figure*}[htb!]
\centering
\includegraphics[width=0.7\linewidth]{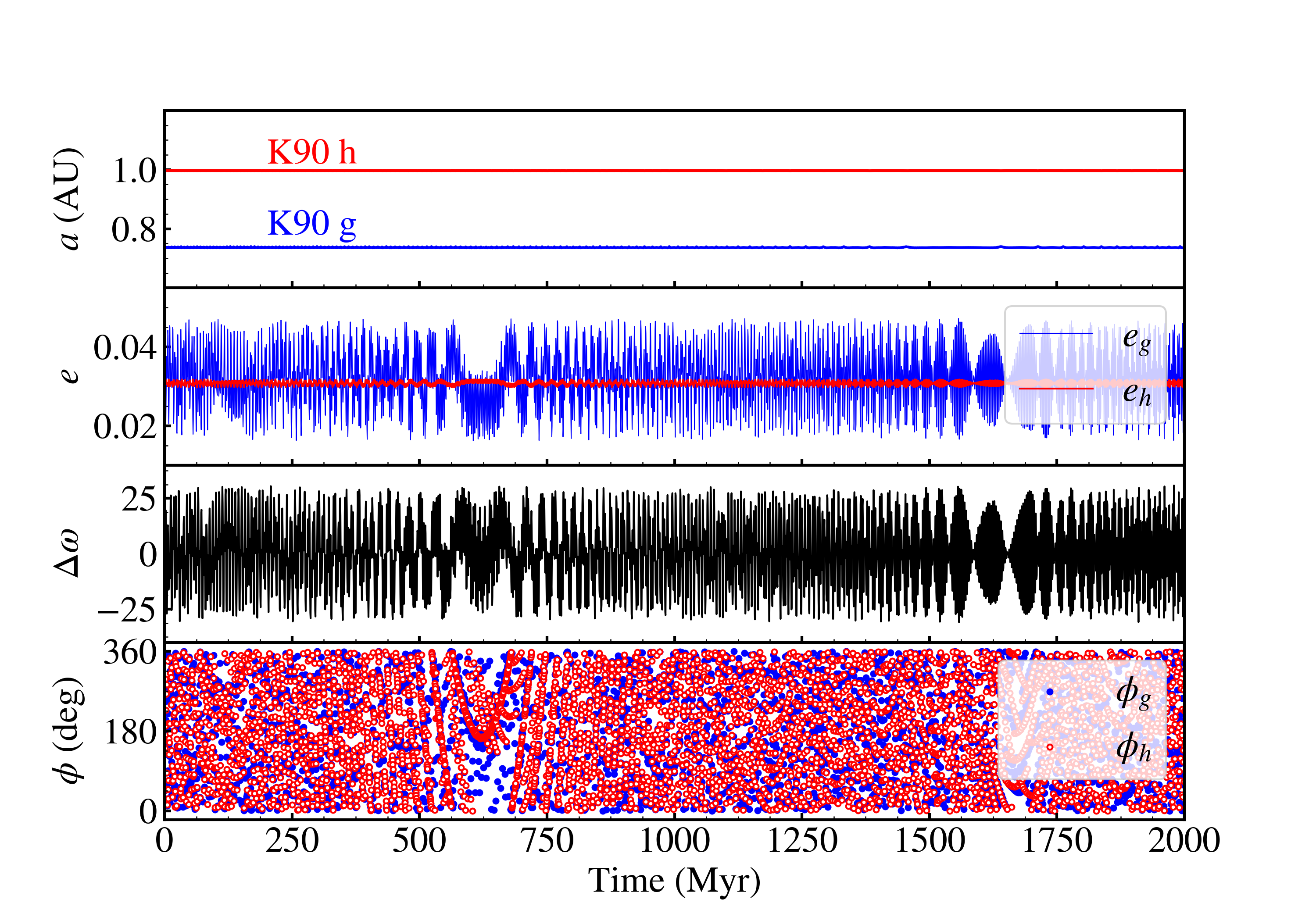}
%includegraphics[width=13cm,height=13cm,keepaspectratio]{longterm.png}
\label{fig:longterm}
\caption{Long term stability integration: This plot shows the evolution of the best two-planet solution over 2 billion years integration. The top panel is the semi-major axes of Kepler 90g and  90h: both are stable over the course of integration. The middle panel shows the eccentricity of 90g (blue), and 90h (red). Due to the stronger gravitational pull of 90h, $\dot{e}_g$ is significantly larger than $\dot{e_h}$. The bottom panel illustrates the evolution of the resonant angles $\phi_{g}=3\lambda_h - 2\lambda_g-\varpi_{g}$ and $\phi_{h}=3\lambda_h - 2\lambda_g-\varpi_{h}$. Both angles span (0,360) degrees, which means that Kepler 90g and 90h are not locked in resonance. However, 90h appears to go in and out of a temporary resonance libration at around 600Myr.}
\end{figure*}

The best solution is close to co-planar (mutual inclination is $-2\pm6^\circ$),  and the arguments of periapsis are aligned ($\Delta \omega$ librates over $\pm 20 ^\circ$). These factors bring the two planet‘s perigees closer to each other and enhance the gravitational pull when they encounter. Dynamical studies show that in co-planar configuration, apsidal resonance can be produced by secular interactions among resonant planets, which also excite the eccentricitie \citep{chiang2002eccentricity,chiang2001apsidal}. The apsidal alignment and relatively high eccentricities of the Kepler 90g and 90h system may be explained by the same mechanism.

\begin{figure}
\centering
\includegraphics[width=\linewidth]{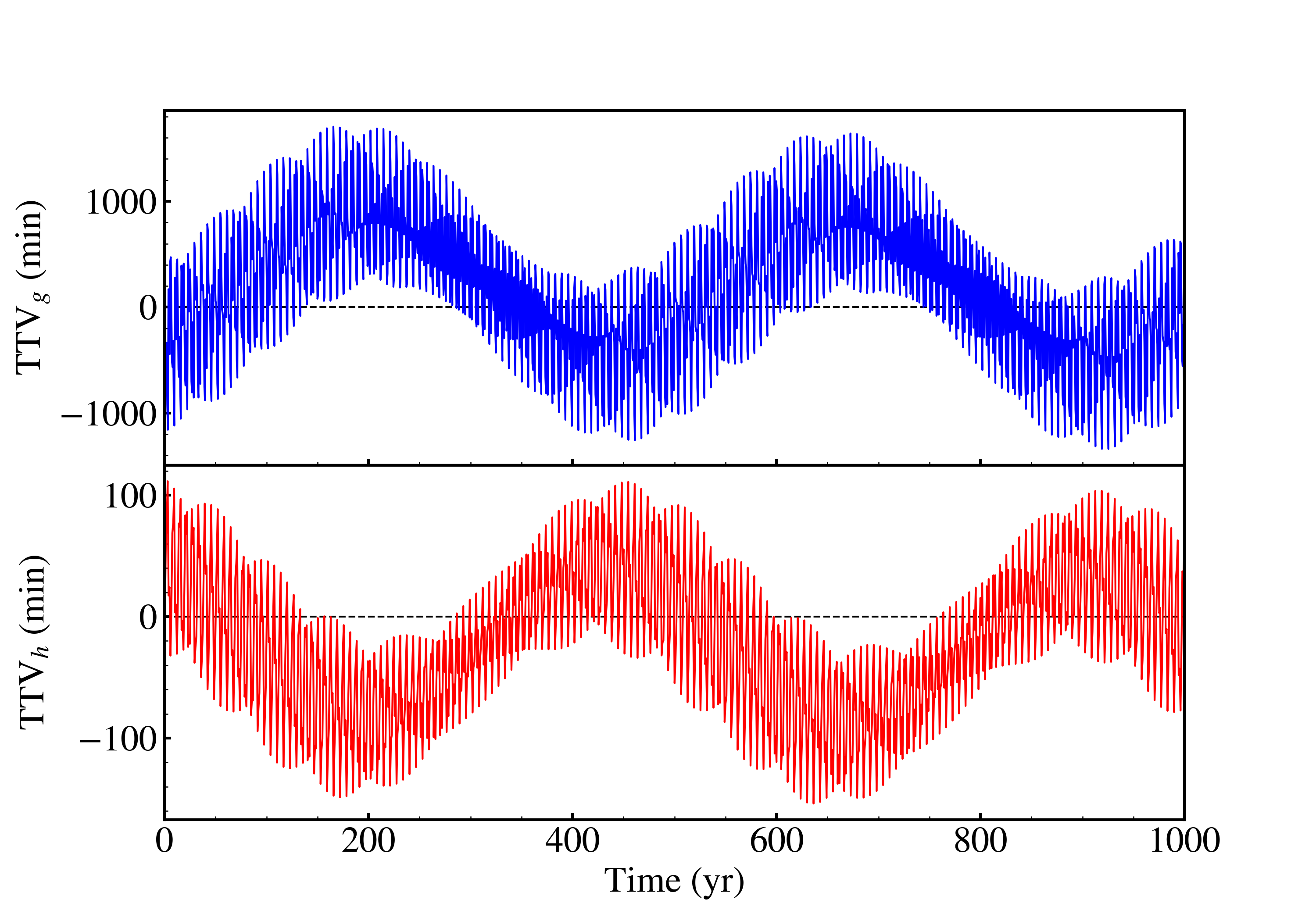} 
\label{fig:longTTV}
\caption{TTV evolution over one thousand years: we show the evolving  TTV signals of Kepler 90g (top) and Kepler 90h (bottom). We use the average period during the integration as the reference point to avoid drifting, so the initial "peak" of Kepler 90g TTV at the epoch starts at -1000 min, instead of 0 min as in figure 1). The TTV patterns are complex, involving many harmonic components. %The secular evolution of TTV signals suggests that the orbital lock does not occur: the system does not come back to the same configuration over several periods.
}
\end{figure}

We monitor the TTV evolution for one thousand years: Figure 3 shows that the TTV is very non-sinusoidal and dominated by secular variations. The orbital stability of Kepler 90g and 90h is not caused by orbital locks, but by the low eccentricity, small mutual inclination, compact orbits, and aligned periapsis. 

\section{Discussion}

While Kepler 90h has mass and density typical of giant gas planets, the best-fit mass of Kepler 90g is $15\pm 0.9 M_{\bigoplus}$, and the radius is $8.1\pm 0.8 R_{\bigoplus}$ \citep{santerne2016sophie}. Combining the two we obtain the apparent 90g density of $0.15\pm0.05\,{\rm g\,cm^{-3}}$, which suggests it is among the lowest density planets found to date, most of which have been found with TTVs in the shorter period planets \citep{masuda2014very,jontof2014kepler,ofir2014independent}. One possible explanation is that it 
has a dusty atmosphere that inflates the observed radius \citep{wang2019dusty}.
Super-puffs typically reside close to the host star, less than 
0.3 Astronomical Units (AU), but in \cite{lee2016breeding} it is argued that super-puffs likely acquire their atmosphere through nebular accretion at around 1 AU, where the disk gas is cooler and less dusty, and then migrate to the inner orbits where they are found. Kepler 90g is at 0.7 AU, relatively close to the expected formation site, with the resonant repulsion and subsequent stable orbit responsible for not migrating inward. 
An alternative explanation is that these planets have large optically-thick rings, which could give an appearance of a larger planet. We have searched for a signature 
of a tilted ring in the transit data \citep{Heising15}, but found no evidence of it. This however does not mean that this explanation is ruled out, as the data do not have sufficient 
sensitivity to distinguish between most of the ring versus no-ring solutions. 

Our analysis suggests that even with just nine transits
between the two planets, a full dynamical analysis of large TTVs can constrain all of the orbital parameters for the two planets, and break the degeneracies present in low TTVs. 
While the low number of transits requires a difficult high dimensional optimization, the solution we found is unique, leading to an unprecedented precision of mass and orbital parameters determination, and enabling dynamical analyses such as the long term evolution of the system. 
A systematic search for TTVs in Kepler data has revealed several candidates with periods above 100 days \citep{Holczer16,Ofir18}, but to date, there has been no systematic search of large TTVs. 
If more systems with large TTVs can be found it would open up a new window into the study of exoplanet formation and evolution models, and a new way to characterize exoplanet demographics.

\acknowledgments

We thank Eugene Chiang, Joshua Winn and 
Daniel Tamayo for constructive discussions and comments. 
This material is based upon work supported by the National Science Foundation under Grant Numbers 1814370 and NSF 1839217, and by NASA under Grant Number 80NSSC18K1274. We acknowledge Ad futura Slovenia for supporting J.R. MSc study at ETH Zürich.


\vspace{5mm}

\software{ttvfast \citep{ttvfast},  
          rebound \citep{rebound}, 
          emcee \citep{emcee}
          }



\appendix

\section{Methods}
\subsection{Data Preparation and Analysis}
We first describe how the TTV and TDV data and its errors are extracted from the Kepler lightcurves. We apply the method developed in \cite{FourierGP}.
We use PDCSAP flux of the Kepler data\footnote{\url{https://exoplanetarchive.ipac.caltech.edu/bulk_data_download/}} processed through the Pre-search data conditioning module \citep{kepler_process}, meaning that long term trends and most of the systematics have already been removed. We use long cadence and, where available, short cadence, light curves with 29.4 and 1 minutes spacings, respectively. 
\par
A light curve is modeled as a sum of the loss of light due to the transit  $U$, stellar variability $s$, and noise $n$, composed of the Gaussian white noise and the non-Gaussian outliers.
The planet $j$ is parametrized by the amplitude of its transits $A_j$ and for each transit, $i$ by its duration $D_{ij}$ and epoch $T_{ij}$. Each transit is calculated by integrating the quadratic limb darkening model for the star's intensity profile over the planet's shadow and over the time spacing interval. Contributions of different transits are then added together.
\par
We model stellar variability as a Fourier Gaussian process. A Fourier transformation introduces a convenient basis $s(\nu) = \mathcal{F} (s(t)) (\nu)$. Two point correlations between the different Fourier components $s(\nu)$ vanish by stationarity, so it suffices to impose a hyperprior power spectrum $P(\nu) = \langle \vert  s(\nu) \vert ^2 \rangle$. 
The power spectrum is learned from the data in an iterative process
where the largest planets are first found and removed and the power spectrum is updated to search smaller planets.
\par
Noise is not correlated but is non-Gaussian: it is composed of the Gaussian central part and a small number of the outliers which deviate significantly from the average. A noise probability distribution can be modeled as a mixture model, whose parameters can be learned by examining residuals averaged over multiple stars.
\begin{equation}
  n(t)= y(t) -  s(t | s(\nu)) - U(t| T, D, A)
\end{equation}
and are distributed according to the noise probability distribution $p(n)$. Parameters can be extracted by minimizing the negative log-likelihood function
\begin{equation}
 \mathcal{L} (T, D, A, s(\nu))= -2\sum_t \ln p(n(t)) + \sum_\nu \left[\frac{|s(\nu)|^2}{P(\nu)}+\ln P(\nu)\right].
\end{equation}
The planet model for Kepler 90g and Kepler 90h depends on $T$, $D$ and $A$, which together contain 20 parameters, while $s(\nu)$ accounts for additional thousands of parameters. Optimization is tractable if we iterate on minimization with respect to the stellar parameters at fixed planet's parameters and vice versa. The final result are the values of the TTV-s, TDV-s and their errors, as well as the latent parameters. The 
procedure is near optimal, and has been shown to give smaller and more reliable errors relative to the 
other existing pipelines \citep{FourierGP}. 
Posterior distribution of the parameters around their optimal values is somewhat non-Gaussian, but we summarize it by a single symmetrized one sigma error to make further analysis tractable.

\subsection{Parameter Space Description and Choice of Priors}

The simplest system we model is that of the 
hosting star and two planets, 90g and 90h. \cite{kepler90} reports the Kepler 90 star mass as $1.2 \pm 0.1 M_{\odot}$.
We fix the stellar mass to 1.2 solar mass in the optimization because TTVs are only sensitive to the mass ratio between the perturber and the hosting star.

In the two-planet scenario, there are 7 parameters for each planet: planet mass, orbital period, inclination, the argument of ascending node, the argument of periapsis, and the mean anomaly at the epoch (HJD 2454833). We fix the argument of ascending node of Kepler 90g to zero, which effectively defines the $x$ axis, so we have 13 free parameters to optimize.

We determine the parameter priors based on the physical knowledge about this system. We anticipate the 90h mass to be much larger than that of 90g since the amplitude of 90g TTV is ~50 times of 90h TTV, which directly reflects the mass ratio between the two planets. In the fit, the 90g mass is allowed to vary from 5 to 50 Earth mass, and the  90h mass range goes from 5 to 500 Earth mass. 

To determine the prior on eccentricities  
we conduct the long-term integration using the \texttt{rebound} package \citep{rebound}, to find the upper limit on eccentricities imposed by the dynamical stability. From a preliminary run, we observe a strong correlation between Kepler 90g and 90h eccentricities, so we assume that they are the same in the high $e$ integration. We first draw a sample from the posterior distribution with $e=0.06$ and $\chi_{TTV+TDV}^2=10.6$. Then, we manually change $e$ and run long-term integration. For high eccentricity orbit ($e>0.3$), the system is unstable in ~$10^5$ years, much shorter than the age of the system. However, for low-eccentricity orbits ($e<0.15$), the stability over millions of years is almost guaranteed, and does not help exclude regions in the parameter space (Figure 4,5).
We thereby restrict the eccentricity of both planets to [0,0.15].

The hard-cutoff for inclinations is set to [89.8,90] degrees. The prior range of the remaining angular arguments are determined in the following manner: the 90g argument of ascending node is fixed to 0 degree such that the initial 90g orbital plane defines the x-axis. The 90h argument of the ascending node, which represents the mutual inclination between the two orbital planes, is allowed to vary from -180 to 180 degrees. The argument of periapsis of both planets is allowed to vary from -180 to 180 degrees since we have no prior knowledge about this parameter. Lastly, the initial guess of the mean anomaly at the epoch is obtained from the first transit time reported by \citep{kepler90}: Kepler 90g Epoch (HJD-2454833) = 147.0364; Kepler 90h Epoch (HJD-2454833) = 140.49631. In the fit we still allow it to vary from 0 to 360 degrees. 
%In the fit, we optimize in 13-dimensional parameter space.

\subsection{Likelihood function}
The likelihood function we use in the optimization consists of two parts: the TTV term and TDV term, 
\begin{equation}
\log{L}(\boldsymbol{p}) = \sum_{\text{j}} \sum_{\text{i}}-\frac{1}{2} \log({2 \pi \sigma^2_{T,ij})}    -\frac{  (\hat{T}_{ij} - T_{ij})^2  }{2 \sigma^2_{T,ij}}                 -\frac{1}{2} \log({2 \pi \sigma^2_{D,ij})}    -\frac{  (\hat{D}_{ij} - D_{ij})^2  }{2 \sigma^2_{D,ij}} .
\end{equation}

Here $T_{ij}$ represents the observed transit times of the $i^{\text{th}}$ transit and the  $j^{\text{th}}$ planet,  $\hat{T}_{ij}$  is the corresponding value predicted by the N body simulation, and $\sigma_{T,ij}$ is the observational error of the transit times. Similarly, $D_{ij}, \hat{D}_{ij} $ and  $\sigma_{D,ij}$ are the observed transit duration, predicted transit duration, and observed duration error of the  $i^{\text{th}}$ transit and the  $j^{\text{th}}$ planet. We have two planets so $j=1,2$. $i$ runs from 1 to 6 for Kepler 90g ($j=1$) and 1 to 3 for Kepler 90h ($j=2$).

We compute  $\hat{T}_{ij}, \hat{D}_{ij}$ given a set of initial parameters using the TTVFast package \citep{ttvfast}. While the recommended step size in the TTVFast is 1/20 of the shortest period, or 10 day in our case, 
we make a rather conservative choice of 0.01 day to ensure accuracy. We explore the consistency of TTVFast by running the integrator at different step sizes ($\Delta t=10^{-5}, 10^{-4}, 10^{-3}...$ day) and monitor the resulting $\chi^2$. We see no observable effect on $\chi^2$ below  $\Delta t=1$ day.

\subsection{Optimization strategy}
Our optimization routine calls the L-BFGS algorithm repetitively through a basin-hopping optimizer to find the planetary parameters that best reproduce the observed signal \citep{wales1997global,zhu1997algorithm}. Due to the high non-linearity, optimization in such a high dimensional parameter space is challenging. To speed up the convergence, we employ two strategies: changing the basis, and fitting parameters group by group. 

In practice, we add the argument of periapsis and the mean anomaly into a single-phase parameter, which significantly decorrelates the two variables. We also observe a strong correlation between $e_{1}, e_{2}$ and $\omega_g, \omega_h$, so we use $\Delta e=e_{h}-e_{g}$,  $\Delta \omega=\omega_{h}-\omega_{g}$ as independent variables in the fit.

A "run" is one call to the basin-hopping optimizer. During a single run, the optimizer takes in an initial guess, a likelihood function, and the bounds of parameters (priors).  To avoid a complex behavior of a high dimensional likelihood function, we divide the thirteen parameters to many groups and only optimize one group at a time, holding the other parameters fixed.

\subsection{Point Search}
Here we describe a typical fitting process.  When we start a new fit, we first make an initial guess $\boldsymbol{p_0}(m_g,P_g,e_g,...,m_h,P_h,e_h,...)$, such as zero eccentricity, aligned orbital planes, etc. 
In the first run, the goal is to first locate an orbit that gives the correct order of magnitude amplitude of TTVs, so we only optimize $m_g,m_h,e_g,e_h$ which determine the overall shape of the orbit. Every time the likelihood function is called, we check if the  $\chi^2$ is lower than $\chi^2_{\text{best}}$. If so, we update  $\chi^2_{\text{best}}$ and save that current parameters as $\boldsymbol{p_{\text{best}}}$ as  $\boldsymbol{p_0}$. If no improvement is made in fifty consecutive calls to the likelihood function, we stop.
After that, we turn to fit the angular arguments which define the spatial configuration of the orbit: We optimize $i_g,i_h,\omega_g,\omega_h,\Omega_h,...$ holding the other parameters fixed. Then, we perform adjustments on the coarsely determined parameters by fitting one planet at a time. This process is to be repeated as many times as necessary. In practice, we optimize all the subgroups sequentially for at least three times to achieve an efficient reduction of $\chi^2$.
The above fitting strategy starts from a point $\boldsymbol{p_0}$ and then ends at a point  $\boldsymbol{p_{\text{best}}} $, and so we call it a "point search" for convenience.

\subsection{Grid Search}
In addition to point search, we also employ a grid search strategy to scan over the parameter space. The goal is to locate all the local minimums such that we are confident about the global minimum we found and its uniqueness. 

Once we have obtained a reasonable solution from the point search, we use it as the new guess  $\boldsymbol{p_0}$. The grid search can be performed on one to three parameters. For example, we start by evaluating the conditional likelihood distribution of two parameters $p_1,p_2$ on a $N\times N$ grid, and fixing other parameters to $\boldsymbol{p_0}$. Then we start a point search at each of the top 30 (lowest $\chi^2$) grid points. It is usually better to stick to the "grids", that is, to hold $p_1,p_2$ fixed during the point search. If we find new local minima that have a lower $\chi^2_\text{min}$ than the previous "center", then we move to the new global minimum and start a new grid search. This process may be repeated until we cannot find other local minima. Since likelihood evaluation is much faster than point search, using dense grids ($N>100$) is usually more efficient than a repetitive point search at the same grid point.

A series of grid searches can find local minimums far from the initial guess. In practice, we put emphasis on three parameters: Kepler 90g periapsis  $\omega_g$, Kepler 90h periapsis $\omega_h$, and the 90h argument of ascending node $\Omega_g$, which defines the mutual inclination between the two planets. 

We chose these three parameters for the following reasons: in the seven independent parameters, four of them are restricted by direct observations. Periods are measured by transit separations; masses can be estimated from the TTV amplitudes; inclinations can be obtained from the transit duration; phases are estimated from the first transits. In addition, long-term integration shows that the eccentricities must be small. Therefore, the only two free parameters are the argument of the ascending node and the argument of periapsis. Since the Kepler 90g argument of ascending node is used to define the $x$ axis, we are left with three parameters to explore.

A significant mutual inclination ($>5^\circ$) is excluded by the first coarse grid search, in which we iterate through all the possible combinations of the three parameters, from -180 to 180 degrees in 5-degree steps. The result suggests that a solution with non-zero mutual inclination has \(\chi^2>10^6\), and does not require further exploration. Then we run a two-parameter grid search with a much higher resolution, which confirmed that we are already at the global minimum.

\subsection{Posterior Distribution}

\begin{figure}[!htp]
\includegraphics[width=0.9\linewidth]{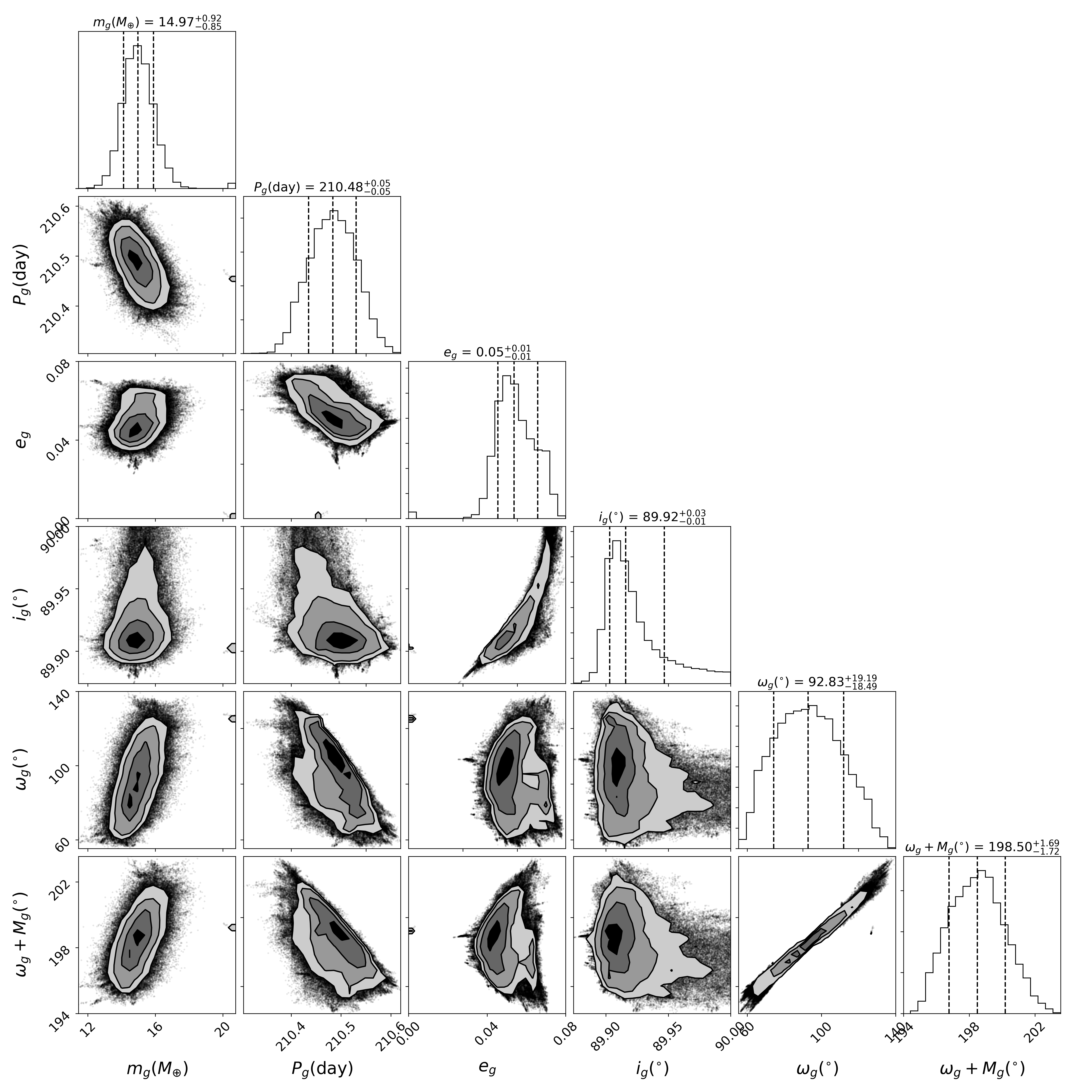} 
\label{fig:90g-posterior}
\caption{The one and two dimensional marginalized posteriors for the 90g parameters.  
}
\end{figure}

\begin{figure}[!htp]
\includegraphics[width=0.9\linewidth]{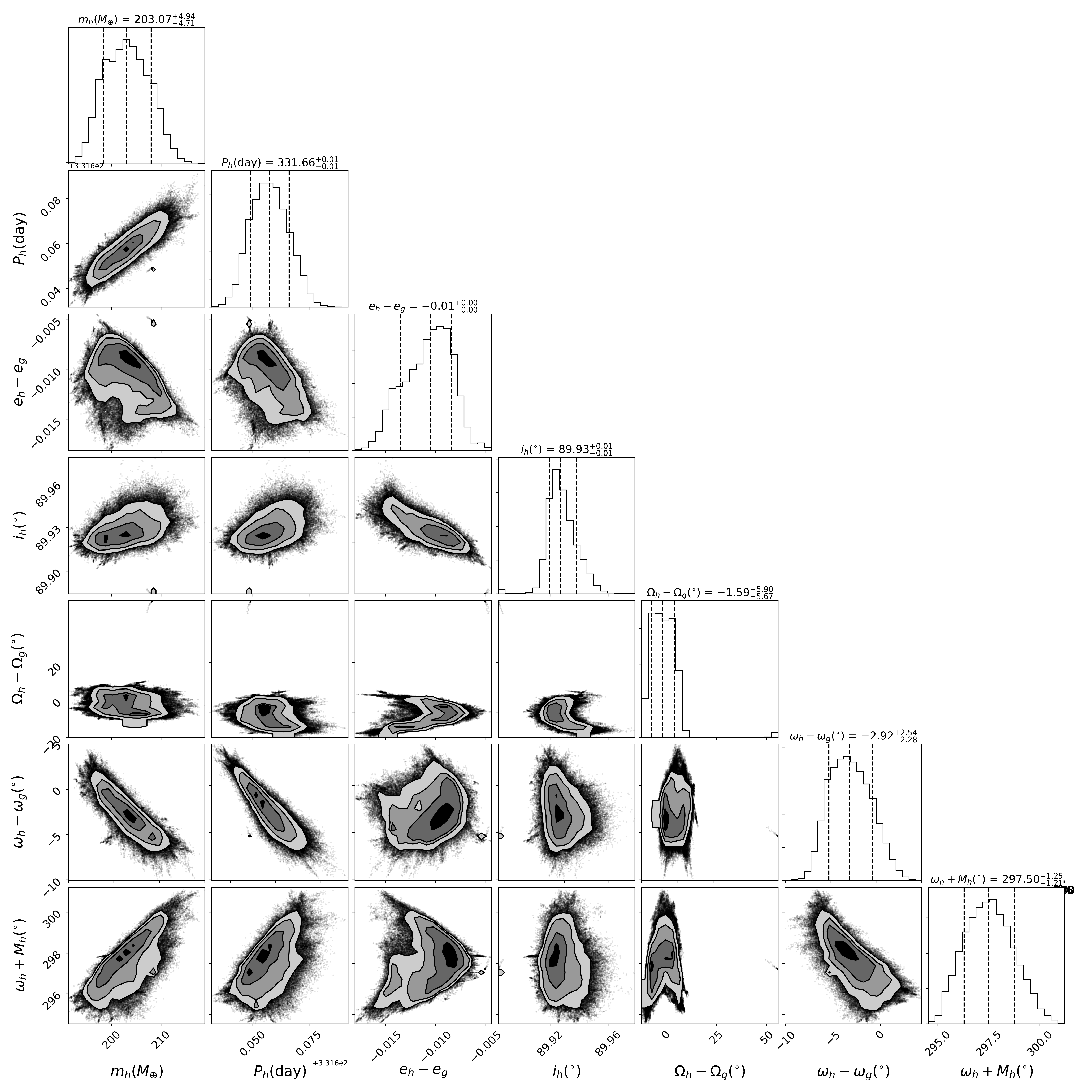}
\label{fig:90h-posterior}
\caption{The one and two dimensional marginalized posteriors for the 90h parameters.  
}
\end{figure}
\newpage

We use \texttt{emcee} \citep{emcee} as the Monte Carlo Markov Chain (MCMC)
sampler.
We configure the \texttt{emcee} sampler with 128 walkers and collect more than $1.6$ million iterations. 
The sampling of the posterior distribution using MCMC is very challenging without the prior knowledge where the peak posterior is.  For this reason, we first identify the global maximum a posteriori (MAP) solution using the optimization described above and start our MCMC around that position. 
Assuming a uniform prior, we do a preliminary run where all the parameters are allowed to vary by 10\% from the best fit MAP solution. The goal is to gather some information around the peak and obtain an estimate of the parameter variance.

Before we start the formal run, we initialize half of the walkers on the multivariate Gaussian distribution that shares the same covariance matrix with the preliminary run. This helps the sampler to identify the interesting regions. The other half of the walkers are also placed on the multivariate Gaussian distribution, but with 5 times increased errors. This allows the sampler to explore better the broader region around the MAP solution..
Finally, we perform boundary checks to make sure that all walkers fall within the prior range. 
The resulting parameter posteriors are shown in Figure 4 and Figure 5.

\subsection{Inner Planets}
We verify that the inner planets (Kepler 90e,f) and the outer planets (Kepler 90g,h) are dynamically decoupled through the following process: we initialize Kepler 90g and 90h to the best fit condition, and add a third hypothetical body on the orbit of Kepler 90f, the closest inner planet, and vary the mass. The TTVs on Kepler 90f due to Kepler 90g and 90h are below $\sim$ 30 minutes and are comparable to the observational error.

When 90f mass is below $1M_{\oplus}$, the perturbations on 90g and 90h orbit are negligible. When the 90f mass is between $1\sim3M_{\oplus}$, we observe a perturbation of a few minutes, formally detectable in the short cadence data. However, such a small change to TTVs can be compensated by a slight modification to the initial conditions, leaving no observable effects to the parameter constraints .

\subsection{Addition of a third body}
It is worth exploring if an alternative 
solutions can be found without resorting to mutual 
perturbations between 90g and 90h. 
We tested alternative hypotheses 
that introduce additional bodies into the system. 
We introduce a new seen or unseen 
body, such as a planet, asteroid, or a Trojan planet, and search over a wide
range of their parameter space in period, phase, mass, eccentricity, 
inclination, ascending node, and periapsis. The best planet solution we find gives 90g TTV residual $\chi^2$ of 20, which is a worse fit to the data and does not eliminate the mutual perturbations between 90g and 90h  as the main origin of TTVs. As a consequence, hypothesis testing with Bayes factor 
shows a large Occam's razor penalty for this solution, due to the large 
prior in the seven-dimensional parameter space of the 
new body.
We also explored 
an exomoon orbiting 90g with three additional parameters, moon period, phase, and mass.
We assume circular orbits aligned with 90g orbit, which only require 3 extra parameters: mass, period, and phase (mean anomaly). We assign the previous 90g parameters to 90g-moon barycenter and require the moon mass to be lower than that of 90g. The lower bound of the moon period is determined by the Roche limit of 90g, which is 0.09 days, and the upper bound is 30 days, at 0.4 Hill Radius. The mean anomaly is allowed to vary from -180 to 180 degrees since we do not have prior knowledge about the moon phase.
We obtain residual $\chi^2=9.51$ for 18 
data points, which is a good fit to the data, but 
the parameters for the two planets are nearly 
unchanged relative to the solution without the 
exomoon, and again the Bayes factor strongly 
disfavors this solution. We thus conclude that the 2 planet solution 
we found is strongly favored and there is no need to add a third body, and there are no found solutions that do not involve 
the mutual perturbations between the two planets as the 
the main origin of observed TTVs. 

\section{Author Contributions}
The project was designed by Y.L, J.R. and U.S. J.R. reduced the TTV data, Y.L. performed the data analysis. The paper was written by Y.L., J.R., and U.S.

\section{Data Availability}
The Kepler stellar flux data are obtained from a public data repository \url{https://exoplanetarchive.ipac.caltech.edu/bulk_data_download/}. The reduced TTVs and TDVs can be found in the supplementary information files. 

\section{Code Availability}
The TTV data reduction code is publicly available in \url{https://github.com/JakobRobnik/Kepler-data-analysis}.


\bibliography{main}{}
\bibliographystyle{aasjournal}



\end{document}